\documentclass[aps,pra,twocolumn,amsmath,amssymb]{revtex4}
\usepackage{bm}
\usepackage{graphicx}

\usepackage{amsfonts,epsfig}
\usepackage{latexsym}
\usepackage{epsfig}
\usepackage{amsmath}
\usepackage{amssymb}
\usepackage{amsfonts}
\usepackage{amsthm}
\usepackage{mathrsfs}
\usepackage{natbib}
\usepackage{color,verbatim,graphics}
\usepackage{psfrag}

\newcommand{\be}{\begin{eqnarray}}
\newcommand{\ee}{\end{eqnarray}}
\newcommand{\ba}{\begin{eqnarray}}
\newcommand{\ea}{\end{eqnarray}}

\begin{document}

\title{Information causality}

\author{Marcin Paw\l owski}
\affiliation{Department of Mathematics, University of Bristol, Bristol BS8 1TW, U.K.}
\author{Valerio Scarani}
\affiliation{Centre for Quantum Technologies, National University of Singapore, 3 Science drive 2, Singapore 117543}
\affiliation{Department of Physics, National University of Singapore, 2 Science Drive 3, Singapore 117542}
%etc.

%\date{\today}

%______________________________________________________________________ ABSTRACT

\begin{abstract}
We review the literature on Information Causality. Since it's for a book, we don't think an abstract will be needed at all, so we have written this one just for the sake of the arXiv.
\end{abstract}

\maketitle

\section{Certain things should not happen}

Like many people working in quantum information science, Bob had spent a few weeks in the Centre for Quantum Technologies in Singapore, collaborating with Alice. Some time after he left, Alice finished preparing ten tutorials for her module on quantum biology. She thought of sharing them with Bob, who was preparing to teach a similar module in his university. However, the latest policies allow only 1Mb attachment per year to an e-mail \footnote{As the reader may expect, this restriction is \textit{not} really implemented in Singapore at the time of writing: we may have to wait for the next generation of managers.}, and each tutorial alone amounts at 1Mb. Alice is in a dilemma: which tutorial will be the best for Bob? It would be much simpler to let Bob choose. But this means that the information about all the tutorials must be made available in Bob's location. How can that happen by sending only a much smaller amount of information?

Alice remembers having shared with Bob, when he was in Singapore, a one-time pad key and even several qubits maximally entangled with hers. Quantum channels can perform tasks that appear incredible to the classically-minded. Can then these shared resources be helpful for this specific task? Alice does not believe it: she knows that shared randomness and entanglement are no-signaling resources. So, she argues, how could they help in sending new information, like the tutorials, which did not even exist at the time of the sharing?

In this text, we show that Alice's argument is wrong: no-signaling resources could in principle solve that task. Her final conclusion is nevertheless correct: the no-signaling resources that exist in our world cannot solve that task. Why? It is probably beyond physics to answer this question. Maybe simply because certain things should not happen?

\section{The context}

\subsection{Defining quantum physics}

\textit{Definire} means to \textit{find the boundary}. In order to define quantum physics, therefore, one can't invoke the ``typically quantum" notions of coherence and entanglement: if anything, these notions fix the boundaries of classical physics. One really needs to go at the quantum \textit{finis terr\ae}. However, all known natural phenomena can be made to fit in the quantum framework. So, are there any boundaries to be found at all?

We leave the question open regarding boundaries in nature. But there are certainly boundaries in the world of physical theories. In quantum \textit{theory}: (i) physical systems must be described by Hilbert spaces, their pure states by one-dimensional projectors, with the rule that orthogonal vectors describe fully distinguishable states; and (ii) the evolution in time must be reversible. As well known by now, pretty much all the formalism stems from these two requirements: a clear boundary, a sharp definition, and a very successful one. However, curiosity is not assuaged: recipes (i) and (ii) define a boundary \textit{with what}? What is there \textit{outside}? How would physics be if (i) and (ii) would not be true?

\subsection{No-signaling is not enough}

\subsubsection{No-signaling as a principle}

It is far from easy to invent decent, consistent answers to the previous questions. Even the anarchical freedom of science fiction has ultimately produced a single creative alternative: \textit{signaling}, in all its possible variations (faster-than-light travel, teleportation of matter between distant locations, etc). No-signaling is certainly a boundary, and a very constraining one at that: just think how tiny is the portion of the universe that the human kind may hope to visit, unless a family of kind wormholes comes to rescue. So let us take this single suggestion seriously: \textit{is no-signaling the physical principle that defines our (quantum) universe?}

Popescu and Rohrlich were the first to raise this question explicitly, and to find a \textit{negative answer} \cite{PR94}. The counter-example uses a simple mathematical object that had been described some years earlier by Rastall \cite{R85}; nowadays it is customarily referred to as \textit{the PR-box} \footnote{In the remainder of this section, we introduce notions and tools that are pretty basic for people working in the field, in order to address a more general readership and also to have a consistent discourse in the text. The reader in need of a more tutorial introduction can refer to Section 5 of: V.~Scarani, \textit{Quantum information: primitive notions and quantum correlations}, in: C.~Miniatura et al.~(eds), \textit{Ultracold Gases and Quantum Information -- Les Houches 2009 session XCI} (Oxford University Press, Oxford, 2011). Preprint available as arXiv:0910.4222.}.

\subsubsection{The PR-box and the CHSH game}

\begin{figure}
\centering
\includegraphics[width = 0.3\textwidth]{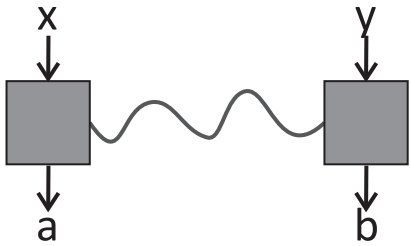}
\caption{The representation of a bipartite no-signaling probability distribution, or ``no-signaling box", used in this text. The wavy line is not meant as a material connection, but only as a reminder of the existence of correlations. The PR-box is defined by $x,y,a,b\in\{0,1\}$, random marginals i.e. $P(a|x)=P(b|y)=\frac{1}{2}$, and perfect correlations satisfying $a\oplus b=xy$.}
\label{figpr}
\end{figure}

The PR-box is a specific bipartite no-signaling probability distribution with both binary input and output (Figure \ref{figpr}). Alice can input a bit $x$ and receives a bit $a$ as output; and similarly Bob can input a bit $y$ and receives a bit $b$ as output. The PR-box is specified by the rule
\ba
P_{PR}(a,b|x,y)&=&\frac{1}{2}\,\delta_{a\oplus b=xy}\,,\label{probpr}
\ea
where the symbol $\oplus$ indicates sum modulo 2. In other words, $a$ and $b$ are always locally random; they are equal in the three cases $(x,y)=(0,0)$, $(0,1)$ and $(1,0)$, while they are different when $(x,y)=(1,1)$.

The PR-box is tailored to violate maximally the Clauser-Horne-Shimony-Holt (CHSH) Bell inequality. For the purpose of this paper, we present this criterion as \textit{the CHSH game}. Alice and Bob are given two binary inputs and must produce, without communication, binary outcomes satisfying (\ref{probpr}). If the inputs are distributed randomly, the probability of success is
\ba
p_{\textrm{CHSH}}&=&\frac{1}{4}\sum_{x,y=0}^1 P(a\oplus b=xy|x,y)\,.
\ea If Alice and Bob are allowed to use only classical shared randomness, their winning probability is bounded as $p_{\textrm{CHSH}}\leq p_{C}=\frac{3}{4}$. If they can share entanglement, their winning probability is increased up to the \textit{Tsirelson bound} \cite{C80}
\ba
p_{\textrm{CHSH}}&\leq& p_{Q}=\frac{2+\sqrt{2}}{4}\,\approx\,85\%
\ea
which is still smaller than one. By construction, the PR-box reaches $p_{\textrm{CHSH}}=1$.

This simple argument proves that no-signaling cannot be the only physical principle that defines our quantum world. At least another constraint is in place, that limits the probability of success of the CHSH game. We can thus rephrase the questions of our curiosity: \textit{given that we live in a world, in which Bell's inequalities are violated, why are they then not violated as much as no-signaling would allow?} Any physical principle (or collection thereof) claiming to come close to a definition of quantum physics should be able to deal with the riddle of the Tsirelson bound.

\subsection{Mathematical framework}

We focus on an operational generalization of quantum kinematics (states and measurement, without dynamics). The measurement process is defined as ``choosing an input and getting an output". The information about the state of the system is contained in the observed probability distributions of the outputs, for each input. Since we focus on bipartite systems, let us fix the notations: the inputs of Alice and Bob are written $x\in{\cal X}$ and $y\in{\cal Y}$, respectively; the outputs (we assume that every input leads to the same number of possible outcomes) are written $a\in{\cal A}$ and $b\in{\cal B}$ respectively. So, for each $x,y$, Alice and Bob can reconstruct the probability distribution $P_{xy}=\big\{P(a,b|x,y)\big|a\in{\cal A},b\in{\cal B}\big\}$. All that Alice and Bob know about the system and the measurements is captured by the \textit{probability point}
\ba
{\cal P}&=&\left\{P_{xy}\big|x\in{\cal X},y\in{\cal Y} \right\}\,.
\ea
\textit{A priori}, each $P_{xy}$ is specified by $|{\cal A}||{\cal B}|-1$ values because of normalization; therefore ${\cal P}$ is generically specified by $|{\cal X}||{\cal Y}|\big(|{\cal A}||{\cal B}|-1\big)$ values.

For the following, it is important to classify probability points as follows:
\begin{itemize}
\item ${\cal P}$ belongs to the \textit{classical set} if it can be written as a convex combination of local deterministic points, i.e. points of the form $P(a,b|x,y)=\delta_{a=f(x)}\delta_{b=g(y)}$. These points are the extremal points of the classical set; since there are finitely many of them, namely $|{\cal A}|^{|{\cal X}|}\,|{\cal B}|^{|{\cal Y}|}$, the classical set is a polytope. In summary, the classical polytope contains all the ${\cal P}$ that can be obtained from ``local (hidden, or not hidden) variables".

\item ${\cal P}$ belongs to the \textit{quantum set} if there exist a state $\rho$ and projectors $\{\Pi_a^x,\Pi_b^y\}$ such that
\ba
P(a,b|x,y)&=&\mathrm{Tr}\left(\rho \Pi_a^x\,\Pi_b^y\right)\,,
\ea where the projectors must satisfy $[\Pi_a^x,\Pi_b^y]=[\Pi_a^x,\Pi_{a'}^x]=[\Pi_b^y,\Pi_{b'}^y]=0$ for all $a,b,x,y$. There is no loss of generality in considering only projective measurements, since the dimensionality of $\rho$ is not restricted. For finite-dimensional Hilbert spaces, these relations between projectors are fulfilled if and only if there is a tensor product representation $\Pi_a^x=\pi_a^x\otimes\openone$ and $\Pi_b^y=\openone\otimes \pi_b^y$ \cite{tsiproblem}.

\item ${\cal P}$ belongs to the \textit{no-signaling set} if $P(a|x,y)=P(a|x)$ and $P(b|x,y)=P(b|y)$ for all $a,b,x,y$. This set is also a polytope. Clearly the classical set is included in the quantum set, which is included in the no-signaling set. Notice also that the no-signaling constraints reduce the number of values required to specify a probability point ${\cal P}$ down to $|{\cal X}||{\cal Y}|\big(|{\cal A}|-1\big)\big(|{\cal B}|-1\big)+ |{\cal X}|\big(|{\cal A}|-1\big)+|{\cal Y}|\big(|{\cal B}|-1\big)$.

\end{itemize}

In this framework, \textit{we are looking for a physical principle, which would single out the quantum set within the no-signaling polytope}.

Before continuing, we want to stress a difference with other operational approaches, in particular with the line of research on axiomatics \cite{axiom}. There, a lot is built on the assumption of tomography: it is supposed that some given ${\cal P}$'s are known to carry all the information the system. This is physically possible if the degrees of freedom under study and the measurements that are being performed on it have been characterized. Here, on the contrary, we work in a \textit{completely black-box scenario}, ultimately the same as in Bell's theorem and in device-independent assessments \cite{di}. In such a scenario, the point ${\cal P}$ can never be claimed to be ``the state", with the idea of complete information that this term conveys. Rather, ${\cal P}$ encodes just the information that can be gathered from the black boxes. This is also one of the reasons why we start out with bipartite systems: in a black-box scenario, the behavior of a single system can always be described in terms of hidden variables.

\section{Information causality: the task}

The statement of ``no-signaling" is the impossibility of a task, namely, sending any amount of information by sampling a bipartite probability distribution. Every device independent principle must have a task (an information processing protocol) and a statement about it. In this section we aim at explaining the choice of the task and the statement of Information Causality. We start by asking the question: in what sense the PR-box is to powerful?

\subsection{The power of the PR-box}

The first device independent principle that put some bounds on the winning probability of the CHSH game was that of nontrivial communication complexity \cite{PR-WvD}. It has been shown that the access to perfect PR-boxes allows two parties to solve any communication complexity problem with the transmission of a single bit. Later this result has been improved in \cite{PR-BRASSARD} where it was shown that the same happens even if the boxes are a little noisy, i.e. they allow for the success probability in the CHSH game greater than $\frac{3+\sqrt{6}}{6}\approx 0.908$. The question whether this principle can be used to derive even stronger limits is still open.

The simple idea behind taking this approach to study nonlocality is that if nothing seems to be wrong with the PR-boxes if the parties are not communicating (and no communication must be the case if we would like to use the no-signaling principle) then maybe there is something wrong with them when the communication takes place. To see why this should be the case let us put ourselves in the place of Bob, the owner of one part of the PR-box. When we choose our setting to be $y=0$ we know that the outcome of our part of the box is going to be equal to the outcome of Alice $b=a$. If we choose $y=1$ instead then we can expect $b=a\oplus x$. We see that we can choose to learn any one of the two independent bits $a$ or $a\oplus x$ by choosing different settings. Granted that these two bits are perfectly random, but their randomness is the same. What we mean by that is that both of them are generated by XORing something deterministic (i.e.~0) or controlled by Alice (i.e.~$x$) with the same random bit $a$. This is important because it allows, by transmitting later only a single bit form Alice to Bob, to erase the randomness in any of the bits that Bob might want to get regardless of his choice.

\begin{figure}
\centering
\includegraphics[width = 0.45\textwidth]{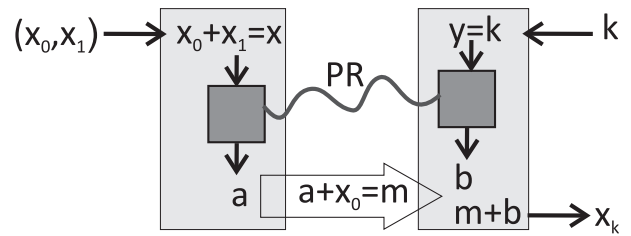}
	\caption{Implementation of perfect oblivious transfer using the PR-box and one bit of communication.}
\label{figot}
\end{figure}

This property of the PR-box has been exploited in \cite{WW} in the context of oblivious transfer (Figure \ref{figot}). Imagine that Alice has two bits $x_0$ and $x_1$. She can send only one bit of classical communication to Bob who is interested in one of the bits (Alice does not know in which). Let the index of the bit that Bob is interested in be $k$. If they have access to a PR-box they can do this. Alice inputs $x=x_0\oplus x_1$ in her part of the box and, after reading $a$, sends the one bit message $m=x_0\oplus a$ to Bob. Bob inputs $y=k$, reads $b$ and computes $C=m\oplus b=x_0\oplus a\oplus b$. It is easy to see that $C=x_k$. Indeed, if $k=0$ then $a=b$ and $C=x_0$; if $k=1$, then $b=a\oplus x$ and $C=x_0\oplus x=x_0\oplus x_0\oplus x_1=x_1$.

Earlier we have promised that this analysis will show us what goes wrong if we consider the protocols with PR-boxes and communication. We are almost there. Look at the situation in the Bob's laboratory when he has already received Alice's message but he has not yet chosen which bit to decode. Considered as a black box his lab now has, in some sense, two bits. True that the extraction of one will destroy the other but, since any can be decoded, they both must be there. But we have transmitted only a single bit and the PR-boxes are supposed to be no-signaling so they cannot be used to transmit the other. Somehow the amount of information that the lab of Bob has is larger than the amount it received. Things like this should not happen.

\subsection{Random Access Codes}

The protocol that we have just described is called (2,1,1) Random Access Code (RAC) \cite{QRAC1}. It allows Alice to encode two bits $x_0$ and $x_1$ into a single bit message $m$ in such a way that Bob can decode any bit he chooses to. The notion generalizes to that of $(N,M,p)$ RAC, which allows Alice to encode $N$ bits into $M$ bit message in such a way that the worst case probability of Bob decoding any of these bits correctly is $p$ \footnote{Earlier we have mentioned that the protocol that we have described is for oblivious transfer. It might puzzle the reader that we are now referring to it as RAC. The difference between these two is that in the oblivious transfer there is one more requirement: Bob after choosing to decode one bit cannot learn anything about the other. In RAC there is no such assumption although in the optimal RACs it is always the case.}. We can talk here as well about the average success probability instead of the worst case since Yao's principle \cite{Yao} applied to RACs allows, with the use of shared randomness, to make these two equal \cite{QRAC2}. There are many different types of RACs with slightly different properties which depend on the resources that we allow to be used. The most important distinction among the known codes lies in what is being communicated (classical bits or qubits).

In the code presented above the bits are decoded correctly as long as the correlations $a=b$ for $y=0$ and $a=b\oplus x$ for $y=1$ are always true. If they occur with probability $p$ then the box can win the CHSH game with this probability and, at the same time, the average success probability of $(2,1,p)$ RAC is also $p$. Therefore, we see that finding a way to bound the success probability of the RAC is equivalent to finding the bound on the probability to win the CHSH game.

\subsection{Task and statement of Information Causality}

\begin{figure}
\centering
\includegraphics[width = 0.45\textwidth]{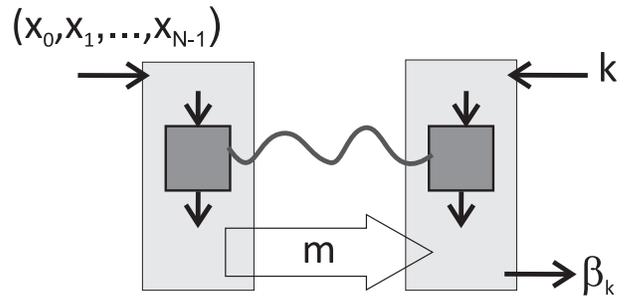}
\caption{The task of Information Causality is the same as the one that defines a Random Access Code. Alice receives $N$ bits, and Bob is asked to guess one of them. Alice is allowed to send a message $m$ that carries $M$ bits, where $M<N$ to avoid trivialities. Moreover, Alice and Bob can share a no-signaling resource --- and in fact, in all this study the goal is to compare the power of such resources. The usual figure of merit is the success probability $p=\sum_{k=0}^{N-1}\textrm{Prob}(\beta_k=x_k|k)$; Information Causality rather quantifies the amount of information that is potentially available in Bob's location.}
\label{figic}
\end{figure}

We are now in a position to define the \textit{task}, to which the principle of Information Causality is going to apply. It is the same as a $(N,M,p)$ Random Access Code, where $N$ and $M$ are classical bits (Figure \ref{figic}). Notice that it does not matter how this information is encoded: when we refer to ``sending the $M$ bit message", it should be understood as a single use of a channel with classical communication capacity $M$.

The \textit{statement} of Information Causality requests that, in the task just defined, \textit{the amount of information potentially available to Bob about Alice's input cannot exceed $M$ bits}. This potentiality is the key to the Information Causality's success. If we would consider only the information that Bob {\it actually} gets, then this principle would be equivalent to no-signaling (indeed, imposing that Bob can actually receive only $M$ bits is equivalent to stating that any additional resource is no-signaling). However, this little tweak makes all the difference as we will see in the next section.

\subsection{The reason for the name}

But before we get to it, we would like to take this opportunity and make a short comment on the choice of the name for our principle. We do this mainly because several people have asked us for the justification of our choice.

Let us reiterate that \textit{Information Causality is about forbidding more information to be potentially available to the receiver than has been sent by the sender}. We hope that expressing our principle in that form makes the choice of the name clearer. Causality is the ability to change something over space-time. In the task we are considering, what gets changed is the information that Bob has about the particular bits of Alice. Before the protocol is run it is, by definition, zero. The cause is the transmission of the message, which increases the information. The statement about the task is that this increase in information is limited. In other words, we are putting a limit on the effect that the cause can have in the terms of information. Hence the name.

\section{Mathematics}

\subsection{The figure of merit}

Now we are ready to present the principle of Information Causality (shortened IC from now on) in its formal version. There are many different measures of information to choose from but in our case the choice is quite obvious. Since the task is about communicating over a channel with a specified classical communication capacity and because Shannon's celebrated single letter formula relates it to mutual information we take this measure. Therefore, the amount of information that Bob can potentially have about the variable $x_i$ of Alice is given by $I(x_i:\beta_i)$ where $\beta_i$ is the random variable that he generates when using his optimal procedure for maximizing the amount of information about this particular $x_i$. The statement of IC is that
\be \label{ic}
\sum_{i=1}^N I(x_i:\beta_i)\leq M.
\ee
Note that we the variables $x_i$ do not have to be binary. We do not make any assumptions about their alphabets. The definition of IC that we have given here is slightly stronger than the one given in the original paper \cite{PPKSWZ09}. There we have assumed that the communication from Alice to Bob is over a noiseless classical channel. This assumption can be lifted and, as we show in the next section, our principle will still hold in the quantum theory.

\subsection{Information Causality holds for quantum no-signaling resources}

IC sounds like a reasonable thing to expect from the universe but so does locality, determinism or the notion of absolute time. Therefore, in the presentation of a new principle, there should always be a proof that it is not violated by nature. Now we present a proof that IC holds in the classical and quantum information theory. We focus on quantum correlations because classical correlations form a subset of quantum correlations.

Let us denote by $\rho_B$ Bob's part of the shared quantum state and $\vec{x}$ the set of all Alice's variables $x_i$. We begin by showing that after receiving the message $\vec{m}$, which was communicated over the channel with the classical communication capacity $M$, from Alice all the classical and quantum information he has does not have more than $M$ bits of information about $\vec{x}$:
\begin{equation}
  I(\vec{x} : \vec{m}, \rho_B) \leq M.
  \label{GNS}
\end{equation}
For the proof we use the chain rule for mutual information, $I(\vec x : \vec{m}, \rho_B) = I(\vec x : \rho_B) + I(\vec x : \vec{m} | \rho_B)$. Since at the beginning of the protocol Bob knows nothing about the variables of Alice $I(\vec x : \rho_B) = 0$,
and the second term $I(\vec x : \vec{m} | \rho_B) = I(\vec x, \rho_B :  \vec{m}) - I(\rho_B :  \vec{m})$
is bounded by $M$ due to the positivity of the mutual information and the fact that $\vec{m}$ is a message sent over the channel with the classical communication capacity $M$.

In the case of independent Alice's input bits condition (\ref{GNS}) limits the information gain about the individual bits as well
because
\begin{equation}
  I(\vec x : \vec{m}, \rho_B) \ge \sum_{i=1}^N I(x_i : \vec{m},\rho_B).
  \label{IND}
\end{equation}
This inequality is also proved using the chain rule. Finally, we observe that Bob's output bit $\beta_i$ is obtained at the end from $\vec{m}$ and $\rho_B$. Hence, the data processing inequality implies $I(x_i:\vec{m},\vec{B})\geq I(x_i:\beta_i)$ which gives us (\ref{ic}).

\subsection{Information-theoretical derivation of the Tsirelson bound}

Here we show that any theory which allows for the violation of the Tsirelson bound violates also IC. To this end we consider a concatenated RAC. Let us explain what we mean by this.

Previously we have presented a code which encodes two classical bits into a single one and gives the average probability of correct decoding equal to the winning probability of the CHSH game. We may think about it as a pair of black boxes. Alice puts two bits into hers and it returns a single bit which she sends to Bob. Bob then puts this message into his box, makes a choice which bit he wants to learn and gets a value which with the probability $p$ is equal to the bit he is interested in. Now imagine that Alice gets four bits instead of two and she is still limited to one bit of communication. She and Bob can construct a RAC for this task with the pairs of the same boxes they used previously with the help of concatenation procedure. It works like this: The parties need three pairs of boxes. Alice puts two of her bits into her first box and the remaining two into the second. The boxes have produced two messages which she does not send to Bob but puts into her third box, instead. It is the output of this final box that she sends to Bob. He inputs it to his box from the third pair and chooses to learn the message generated by the first or the second box of Alice. He inputs this message into one of his other boxes - the one paired with the box of Alice that generated this message, and then he can retrieve the bit. The overall success probability is now $p^2+(1-p)^2$ if the success probability for each pair of boxes is $p$.

The generalization of this procedure is quite straightforward. If the parties use $n$ levels of concatenation (using just a single pair of boxes corresponds to $n=1$) they can encode $2^{n}$ bits using $2^n-1$ pairs of boxes. The overall success probability of decoding the desired bit correctly is $p_n=\frac{1+E^2}{2}$, where $E$ is the bias of the probability $p$ (i.e. $p=\frac{1+E}{2}$).

If $\beta_i$ is Bob's best guess of $x_i$ and they are equal with the probability $p_n$ then $I(x_i:\beta_i)=1-h(p_n)$, where $h(.)$ is Shannon's binary entropy. By expanding it into the Taylor series one gets that
\be
1-h\left(\frac{1+E^n}{2} \right)\geq \frac{E^{2n}}{2\ln 2}.
\ee
Since only one bit has been communicated, IC implies
\be
1\geq \sum_{i=1}^{2^n}I(x_i:\beta_i)\geq 2^n\frac{ E^{2n}}{2\ln 2}=\frac{1}{2\ln 2} \left(2E^2\right)^n
\ee
for any $n$. This is going to be true only if $2E^2\leq1$ or, equivalently, $E\leq \frac{1}{\sqrt{2}}$. This puts a bound on the winning probability of the CHSH game $p\leq \frac{1}{2}\left(1+\frac{1}{\sqrt{2}} \right)$ which is exactly the Tsirelson bound.

Quite straightforward generalization of this method can be employed if the probabilities of guessing different bits are different. In \cite{EARAC} it has been used to derive the bound on the efficiency of the RAC's
\be
\sum_{i=1}^N E_i^2\leq 1,
\ee
where $E_i$ is the bias of the guessing probability for the $i$'th bit.

\subsection{Entropic approach}

In order to prove that IC holds in quantum mechanics we have used the properties of mutual information. This means that something must go wrong with entropy measures for superstrong nonlocal boxes, as indeed was discussed shortly after the first IC paper \cite{entropies}. In the latest development \cite{AS11}, it has been shown that all the properties necessary for the derivation of IC are consequences of only two conditions. This means that even if we choose a measure of information different than the mutual information, the objects exhibiting more nonlocality than the quantum theory allows will violate at least one of these conditions.

The conditions proposed in \cite{AS11} are for the entropies $H(.)$. The information that object $A$ has about $B$ can be defined as for the von Neumann entropies as $I(A:B)=H(A)+H(B)-H(A,B)$. The first of the conditions is consistency: if $A$ is a classical random variable, then $H(X)$ is equal to the Shannon entropy of $X$. The second is evolution with an ancilla: for any two systems $A$ and $B$, whenever a transformation is performed on $B$ alone, one must have
$\Delta H(A,B)\geq \Delta H(B)$. It can be understood as saying that local transformations can only destroy correlations not create them.

Since the consistency condition is nothing more than the normalization of the entropy, it must be the second one which is violated by the superstrong nonlocality. This provides another characterization of what is wrong with no-signaling theories that violate Tsirelson bound: even though they cannot instantaneously send information at a distance, they can create correlations which is just as unacceptable.

Recently a slightly generalized version of IC has been proposed \cite{BG11}. It keeps all its reasonable appeal and leads to entropic inequalities that are strictly stronger than in the original version. Recall that the reasoning that lead us to stating IC included two steps. In the first step, we argued that if the Bob's part of the system together with the message are treated as a single black box, then the information it has about the settings of Alice cannot exceed the classical communication capacity of the channel. If we associate random variable $e$ with this black box we can express this statement formally as
\be \label{w0}
H(\vec{m})\geq I(\vec{x}:e).
\ee
In the second step, we argued that the random variable $\beta_i$ is obtained locally from $e$, therefore the data processing inequality implies
\be \label{wi}
\forall_i \quad H(x_i|\beta_i)\geq H(x_i|e).
\ee
If we sum up all the inequalities (\ref{w0}) and (\ref{wi}) and use the subadditivity of the entropy we obtain
\be
H(\vec{m})+\sum_i H(a_i|\beta_i)\geq H(\vec{x})\,,
\ee
which is equivalent to (\ref{ic}) in the case when the $x_i$ are independent. But nothing forces us to sum up all the terms with the same weight. In fact, we can use a different one for each of the inequalities and get that, for all $w_i\geq 0$ and every $p(e|\vec{x})$, it holds
\be
w_0H(\vec{m})+\sum_i w_i H(a_i|\beta_i)\geq w_0 I(\vec{x}:e)+\sum_i w_i H(a_i|e),
\ee
which is strictly stronger than the original IC. It remains to be seen if this new version of the principle leads to tighter bounds on what is possible in our world and what is not.

\section{(Un?)expected complexity}

The fact that IC solves the riddle of the Tsirelson bound has been considered as a remarkable success. But of course, the ultimate goal is far more ambitious: is IC \textit{the} physical principle that defines our quantum universe? In other words, does IC define exactly the quantum set within the no-signaling polytope, in any scenario? In the following, we refer to this scientific quest as to \textit{the IC program}.

Several subsequent studies have witnessed partial success and lead to a wealth of unanswered questions --- which are of course also an asset for research, at least as long as their complexity does not suffocate the driving motivation. In this last section, we review the status of the IC program.

\subsection{Non-isotropic correlations}
\label{noniso}

The recovery of the Tsirelson bound proves that IC defines the quantum set if one considers the single-parameter family of ``isotropic correlations", that is, the probability points that can be written as a convex combination of the PR-box and the white noise. In the first extension of the basic result, the authors considered whether IC defines the whole quantum set in the CHSH scenario \cite{ABPS09}. The conclusion is that we don't know yet. Specifically, the paper focused on two-parameter families (recall that the no-signaling polytope lives in an eight-dimensional space). The violation of IC is assessed using the same explicit protocol described above, which is not guaranteed to be optimal \textit{a priori}. For some families, IC is found to be violated as soon as one leaves the quantum set; in other cases, a finite gap is left. Similar results have been obtained by studying the probability points that admit a Hardy's paradox \cite{indians}.

Adopting an optimistic view on the IC program, one may surmise that the gap is only due to the specific protocol using concatenated RAC. Indeed, a subsequent paper showed that this protocol is provably not optimal for some points \cite{ABLPSV09}. Indeed, some points, which do not exhibit a violation of IC under that protocol, can be ``distilled" to points which do violate IC under the same protocol. In other words, if the process of ``distillation" is added to the protocol, the gap shrinks. However, it is not yet fully closed. Notice that, apart from the fact itself of belonging to the quantum set, we know don't know any sufficient condition for IC to be respected \footnote{A sufficient condition for IC to hold has been given \cite{taiwan}, but for a fixed protocol (how to use the no-signaling resource, coding of the signal bit etc.); it is therefore of limited scope.}.

The scary part of it all comes when one realizes that we are still speaking of the elementary CHSH scenario: two parties, two inputs and two outputs! Quantum physics is certainly more than that. What can one say for more general scenarios?

\subsection{Comparison with ``macroscopic locality"}

The first natural generalization consists in keeping the bipartite scenario and enlarging the alphabets of the inputs and/or the outputs of the no-signaling resource. Obviously, this can in principle be done by keeping the task as a RAC involving bits. For simplicity, though, the only larger-alphabet study published so far \cite{CSS10} generalized also the task to a RAC in which Alice receives $N$ classical dits and send $M=1$ classical dit to Bob. The underlying no-signaling resources are such that $|{\cal X}|=|{\cal A}|=|{\cal B}|=d$, while $|{\cal Y}|=2$.

The main result of this paper is the observation that IC comes closer to defining the quantum set than does \textit{macroscopic locality} (ML). The latter is another criterion proposed with a similar scope \cite{NW09}. It basically says that, in an experiment with many independent sources, the coarse-grained statistics should not violate Bell's inequalities. For instance, imagine a down-conversion experiment in which one would not be able to count photons and had to rely on proportional counting: then the observed currents and their fluctuations could be compatible with a classical source.

The correlations that satisfy ML have been characterized completely: they form a set which is close, but not identical, to the quantum set. Therefore, it is a necessary condition for the IC program to succeed, that IC can rule out more correlations than ML does. Reference \cite{CSS10} provides examples of correlations for which it is indeed the case.

\subsection{IC and multi-partite correlations}

Complexity is further increased if one moves from bipartite to multipartite situations. Even in the simplest tripartite scenario (two inputs and two outputs per party), the structure of the no-signaling polytope is appalling \cite{PBS11}.

One can certainly take multipartite boxes and use them as underlying no-signaling resource in a bipartite scenario: for instance, in the tripartite case, Alice may hold two of the input-output ports and even wire them together, while Bob keeps the third port. This has been tried, and the result is somehow expected: bipartite IC is powerful enough rule out many examples of non-quantum points \cite{XR11}, but not all. In fact, two different examples have been reported of tripartite probability points, which are definitely not quantum but which exhibit classical behavior in any bipartite scenario
\cite{GWAN11,YCATS11}. Therefore, in order to pursue the IC program, one of the most urgent tasks consists in finding a natural generalization of the IC task to more parties; at the moment of writing, the unpublished attempts we are aware of have not lead to anything interesting.

\section{Conclusion}

Formulated just two years ago, Information Causality has immediately attracted the attention of the scientific community. The reason for this success may be purely sociological: the idea that physics may be defined in terms of information processing has been lingering for many years and IC came to fill in the expectation. But we prefer to think in more ``objective" terms: as we were trying to argue all along this text, IC is a very sensible thing to assume about the universe.

Improvement on the initial study have proved technically challenging: often restricted to extremely specific examples, they have nevertheless provided interesting information about the power of the notion of IC and unraveled some of its complex features. A few more of these specific studies will certainly be welcome; but if the IC program has to succeed, one will have to find a much more comprehensive approach. It is our sincere wish that this short review be outdated soon.

\end{document}